\documentclass[twocolumn,preprintnumbers,amsmath,amssymb]{revtex4}

\usepackage{graphicx}
\usepackage{bm}

\begin{document}

\preprint{USTC-ICTS-08-09}

\title{Eternal Inflation: Prohibited by Quantum Gravity?}

\author{Yi Wang}

\email{wangyi@itp.ac.cn}
\affiliation{%
Institute of Theoretical Physics, CAS, Beijing 100080, P.R.China\\
Interdisciplinary Center of Theoretical Studies, USTC, Hefei, Anhui
230026, P.R.China
}%

\date{\today}

\begin{abstract}
We investigate how eternal inflation is affected by quantum gravity
effects. We consider general features of quantum gravity, such as
renormalizability, complementarity, minimal length, definition of
observables, and weak gravity conjecture. We also consider
phenomenological models such as ghost inflation, non-commutative
inflation, brane inflation, k-inflation and resonant tunneling. We
find that all these features and models do not support eternal
inflation. These evidences show hints that eternal inflation is
prohibited by quantum gravity.
\end{abstract}

\maketitle

\section{Introduction}

It has been widely accepted that the observable stage of inflation
provides the initial condition for a flat universe, as well as the
cosmic microwave background (CMB) and the large scale structure
(LSS). Based on this, it is natural to ask what provides the initial
condition for the observable inflation, or in other words, where our
universe comes from. This selection of original universe program
\cite{Firouzjahi:2004mx, Sarangi:2005cs} is hopeful to answer some
profound questions in physics, such as to determine the fundamental
constants and matter contents of nature.

One possible mechanism to provide the initial condition for the
observable inflation is eternal inflation. Eternal inflation
indicates that inflation in the universe is eternal to the future,
and we live in a locally thermalized bubble embedded in the
inflating background.

There are typically two ways to realize eternal inflation, namely,
the slow roll eternal inflation \cite{Linde:2005ht, Goncharov:1987ir} and the false vacuum eternal
inflation \cite{Guth:1982pn}. The slow roll eternal inflation originates from the
comparison between the quantum fluctuation and the classical motion
of the inflaton per Hubble time. If the quantum fluctuation is
comparable with or larger than the classical motion, inflation runs
into a self-reproducing process and become eternal. The false vacuum
eternal inflation originates from a smaller decay rate of the false
vacuum compared with the inflationary Hubble constant. In this case,
the physical spatial volume during inflation increases and inflation
becomes eternal. If the vacuum structure of the real world is very
complicated, such as the string landscape, then eternal inflation
can happen as a combination of these two types.

Eternal inflation is inevitable semi-classically if the inflaton
potential satisfies the conditions for eternal inflation. However,
as we will discuss in the following sections, eternal inflation
becomes problematic when quantum gravity effects are taken into
consideration. Whether eternal inflation happens or not has deep
implications in cosmology, deciding whether the best information for
the initial condition of our universe is probabilistic or
deterministic.

If eternal inflation happens, it is widely believed that the
deterministic initial condition for the
creation of our universe is diluted. In the best case, if there are
some exponentially favored vacua, we can
determine the initial condition by constructing measures
for eternal inflation. Otherwise, it may be hopeless to determine the initial
condition for our universe.

However, if eternal inflation is prohibited in the quantum gravity
level, then it becomes possible that the initial condition for our
universe can be calculated from the first principle. In this case,
the information for creation of the universe is available, and can
be verified by experiments.

In this paper, we consider how eternal inflation is affected by
quantum gravity effects, from both general arguments and
phenomenological models. In Section 2, we argue that the prohibition of eternal
inflation is rather general in quantum gravity. We conclude
in Section 3.

\section{Quantum gravity phenomenology}

As we shall discuss, quantum gravity is cried for during eternal
inflation. But unfortunately, we do not have a full quantum gravity
theory for cosmology so far. In this section, we shall discuss some
general features and phenomenological models of quantum gravity.

\subsection{General Arguments}

One of the key problems in quantum gravity is the non-renormalizable
nature of gravity. In order to have a renormalizable or finite
theory for gravity, one need to suppress the quantum fluctuations in
the high energy regime. On the other hand, the slow roll eternal
inflation needs large quantum fluctuations. So it is likely for
quantum theory effects to kill slow roll eternal inflation. One
explicit example of this general argument is shown in Subsection D
of this section.

It is well known that it is very difficult to construct a measure
for eternal inflation. Two classes of measures are considered in the
literature, namely, the global \cite{Garriga:2005av, Linde:2007nm}
and local \cite{Bousso:2006ev, Bousso:2006ge, Li:2007uc} measures.
However, regardless of technical difficulties such as divergences or
gauge dependence, both the global and local measures suffer problems
of the nature of quantum gravity.

Global measures are weighted by the spatial volume or the number of
the bubbles of a given kind. These attempts to describe spatial
regions divided by future event horizon, and violates directly the
cosmic complementary principle. So global measures suffer quantum
gravity problems globally \cite{Linde:2007nm}.

Local measures are weighted by how long or how many times a given
vacuum is accessed by an eternal co-moving observer. However, if
inflation has happened for a sufficiently long time, and we track
different co-moving observers back in time, these co-moving
observers run into a single Planck volume. Thus it does not make
sense to distinguish these observers. So local measures suffer
quantum gravity problems locally \cite{Bousso:2006ge, Li:2007uc}.

As we have seen, none of these measures are self-consistent in
quantum gravity. Keeping this problem in mind, it is natural to take
one step back to conjecture that this inconsistency is evidence for
the absence of eternal inflation.

Another hint for the prohibition of eternal inflation is the
difficulty in defining the observables in the asymptotically de
Sitter space. One solution for this problem is to embed the de
Sitter inflating patch into a anti-de Sitter background. However, if
eternal inflation takes place, then the causal structure of the
spacetime is changed and the above picture no longer applies \cite{ArkaniHamed:2008ym}.

Eternal inflation also runs into problems of Boltzmann brains. As
discussed in \cite{Page:2006dt}, if we are typical observers, and the
universe eternally inflates, then we should find ourselves to be
Boltzmann brains instead of human observers. The simplest solution to
this problem is that eternal inflation can not happen. For other
possible solutions, see \cite{Li:2007dh, Gott:2008ii}.

\subsection{Weak Gravity Conjecture (WGC)}

It is conjectured in \cite{ArkaniHamed:2006dz} that gravity should
be the weakest force in a self-consistent quantum gravity theory.
This conjecture leads to a UV cut-off for the effective field theory
of the inflaton. In \cite{Huang:2007zt}, we have shown that eternal
inflation is prohibited by WGC for $m^2\varphi^2$ and
$\lambda\varphi^4$ single field  assisted inflation. Take the
$n$-field $\lambda\varphi^4$ potential for example, by comparing the
interaction strength for the inflaton self coupling with the
gravitational coupling, WGC gives a cut-off for the Hubble parameter
\begin{equation}\label{WGC1}
  H<\Lambda=\sqrt{\lambda}M_p~.
\end{equation}

In order for inflation to take place, the kinetic energy for inflaton
should be smaller than the potential energy,
\begin{equation}\label{WGC2}
  n(\partial_\mu\varphi)^2\sim nH^4<V~.
\end{equation}
The eternal inflation condition takes the form
\begin{equation}\label{WGC3}
  \sqrt{n}\delta_q\varphi >n\delta_c\varphi~,
\end{equation}
where $\delta_q\varphi\simeq H/(2\pi)$ is the quantum fluctuation of
each inflaton, and $\delta_c\varphi$ is the classical motion of each
inflaton during one Hubble time.

One can show that (\ref{WGC1}), (\ref{WGC2}) and (\ref{WGC3}) can
not be satisfied simultaneously when $\lambda<1$ and $n\geq 1$. So
this kind of slow roll eternal inflation is prohibited by WGC.
Similarly, we can prove the same result for the $m^2\varphi^2$
potential.

\subsection{Ghost Inflation}

Ghost inflation \cite{ArkaniHamed:2003uz} is proposed as an IR
modification of gravity. In this subsection, we shall show that slow
roll eternal inflation is absent in ghost inflation. In ghost
inflation, the inflaton background moves with a constant velocity
$\langle\dot\varphi\rangle=M^2$. The perturbation $\pi$ around the
background is defined as $\varphi=M^2t+\pi$, with Lagrangian
\begin{equation}
  S=\int d^4 x
  \frac{1}{2}\dot\pi^2-\frac{\alpha^2}{2M^2}(\nabla^2\pi)^2-\frac{\beta}{2
  M^2}\dot\pi(\nabla\pi)^2+\cdots~,
\end{equation}
where $\alpha$ and $\beta$ are order one constants. The condition for
the effective field theory to hold is
\begin{equation}\label{ghosteffective}
  H\ll m~.
\end{equation}

In \cite{ArkaniHamed:2003uz}, it is shown by scaling arguments that
the quantum and classical fluctuations take the form
\begin{equation}
  \delta_q\varphi\sim(HM^3)^{1/4}~,~~~\delta_c\varphi\sim \dot\varphi/H\simeq M^2/H~.
\end{equation}
So the eternal inflation condition $\delta_q\varphi>\delta_c\varphi$
takes the form $H>M$. This directly violates (\ref{ghosteffective}),
thus slow roll eternal inflation is prohibited.

\cite{ArkaniHamed:2003uz} also discussed two corrections for the above amplitude
for quantum fluctuations, one from the parameter $\alpha$ and the
other from tilting the potential. However, on condition that high
order terms are important, $\alpha$ should not be much smaller than
one, so the $\alpha$ correction does not change the result.

For
tilting the potential, it can also be shown that eternal inflation is
absent. In order to stay within the effective field theory, the
classical motion of inflaton remains almost the same. When the
potential for the perturbation satisfies
$-V'>3H^2M$, the quantum fluctuation is significantly
modified \cite{ArkaniHamed:2003uz}. However, it can be shown that the new expression for quantum
fluctuation also prohibits the slow roll eternal inflation.

So we conclude that after the corrections are taken into
consideration, ghost inflation still can not be eternal.

\subsection{Spacetime Non-Commutativity}

In this section, we consider the spacetime non-commutativity inspired
by string theory \cite{Li:1996rp}. The spacetime non-commutativity takes the form \cite{Brandenberger:2002nq}
\begin{equation}
  [t_{\rm phys},x_{\rm phys}]=i M_N^{-2}~,
\end{equation}
where $M_N$ is the non-commutative scale.

It is shown in \cite{Cai:2007et} that when the non-commutative effect is
strong, the inflaton quantum fluctuation per Hubble time takes the
form
\begin{equation}
  \delta_q\varphi\simeq \frac{1}{2\pi}\frac{M_N^2}{H}~.
\end{equation}

Comparing this result with the classical motion $\delta_c\varphi$,
we conclude that for $\lambda M_p^{4-p}\varphi^p$ potential ($p \geq
2$), eternal inflation will never happen when
\begin{equation}
M_N<\left(p^p \lambda\right)^{\frac{1}{p+2}}M_p~.
\end{equation}
For example, for the $m^2\varphi^2$ potential, eternal inflation will
never happen when
\begin{equation}
  M_N<\sqrt{m M_p}~.
\end{equation}
Plugging in the experimental value $m\sim 10^{-6}$, we find that
when $M_N< 10^{-3}M_p$, eternal inflation can not happen.

So we conclude that the slow roll eternal inflation is prohibited by
a low scale spacetime non-commutativity. It is also interesting to
investigate the possibility for the false vacuum eternal inflation.
We find \cite{Cai:2007bw} that false vacuum eternal inflation with
Hawking-Moss tunneling can happen with non-commutativity. This is
not surprising, because the probability for Hawking-Moss tunneling
is suppressed when quantum fluctuation is suppressed. It should be
interesting to investigate the CDL tunneling with non-commutativity.
We hope we can address this issue in future work.

\subsection{Power Counting for Chaotic Inflation}

Another piece of evidence against eternal inflation originating from
power counting and validity for effective field theory is observed
in \cite{Lyth:1998xn}. we can write the power expansion for the
inflaton potential as
\begin{equation}
  V=V_0 \pm
  \frac{1}{2}m^2\varphi^2+M\varphi^3+\frac{\lambda}{4}\varphi^4+\sum_{d=5}^{\infty}\lambda_d
  \Lambda_{\rm UV}^{4-d}\varphi^d~.
\end{equation}

The terms in the summation ($d\geq 5$) can usually be neglected because they
are suppressed by the cut-off of the effective field theory
$\lambda_{\rm UV}$. However, in the chaotic inflation, $\varphi>M_p$,
so the power expansion is not under control.

One can always shift $\varphi$ to make the above expansion work.
However, in the chaotic inflation, the variation of the inflaton
$\Delta\varphi\equiv\varphi_f-\varphi_i>M_p$. So we still can not
get the effective field theory work along the whole inflation
trajectory. As the most studied slow roll eternal inflation models
are within the chaotic inflation framework, this problem for chaotic
inflation also serves as a problem for slow roll eternal inflation.
Similar conjecture that the inflaton trajectory should not be longer
than $M_p$ is also proposed in \cite{Ooguri:2006in, Huang:2007qz}.
Finally, one should note that there are also arguments against this
power counting reasoning, see, e.g. \cite{Linde:2007fr}.

\subsection{Inflation Models in String Theory}

As is well known, it is very difficult to construct large field
inflation models in string theory. This difficulty makes slow roll
eternal inflation problematic in string theory. For example, in
\cite{Chen:2006hs}, it is shown that brane inflation can not be slow
roll eternal inflation. (However, in \cite{Tolley:2008na}, it is
reported that by adding higher order polynomial corrections to the
potential, eternal inflation can take place in brane inflation.) And
it is noticed in \cite{Leblond:2008gg} that when the sound speed of
the inflaton perturbation has deviation from unity by
$1-c_s^2>\epsilon-2\dot c_s/(Hc_s)$, then on condition that
inflation happens in the perturbative regime, one obtains
\begin{equation}
  c_s^4 >\frac{H^2}{M_p^2\epsilon c_s}\simeq P_{\zeta}~,
\end{equation}
where $P_{\zeta}$ is the power spectrum of the curvature perturbation
on uniform density slice. So when $c_s<1$ (as indeed the case in brane
inflation and most k-inflation models), we have $P_\zeta<1$. This
leads to $\delta_q\varphi<\delta_c\varphi$, so eternal inflation is
prohibited by the small sound speed effect.

Another class of string inflation models is modular inflation. The
existence of eternal inflation in modular inflation is in debate. On
one hand, eternal inflation can happen in some effective field
theory models derived from the reshaping of the compactified
dimensions \cite{racetrack}. On the other hand, some people insist
that no realization of slow roll eternal inflation has been found in
string theory so far \cite{ArkaniHamed:2008ym}.


\subsection{Rapid  Tunneling in the Landscape}

Most of the above discussions are related to the slow roll eternal
inflation. In this subsection, we review briefly the rapid tunneling
in the string landscape \cite{HenryTye:2006tg, Tye:2007ja, Huang:2008jr}, which may prohibit the false vacuum eternal inflation.

One mechanism for the rapid tunneling is resonant tunneling.
Resonant tunneling is a well known effect in quantum mechanics.
Recently, resonant tunneling in quantum field theory is investigated
\cite{Saffin:2008vi}. It is shown in \cite{HenryTye:2006tg,
Tye:2007ja} that the tunneling probability in the landscape with
resonant tunneling takes the form
\begin{equation}
  \Gamma\sim n^d\Gamma_0~,
\end{equation}
where $d$ is the dimension of the landscape, $\Gamma_0$ is the
original tunneling rate without resonant tunneling, and $n$ is the
number of effective steps of tunneling which is not much suppressed
compared with $\Gamma_0$. It is noticed in \cite{Feldstein:2006hm,
Chen:2007gd} that if the CMB power spectrum is produced by a chain
of resonant tunneling, then the tunneling rate should satisfy
$\Gamma \gg H$ during observable inflation. As observed in
\cite{Tye:2007ja}, the tunneling rate increases with inflationary
energy scale, so it is natural that $\Gamma > H$ also holds for
higher energy scale. Then the false vacuum eternal inflation can not
happen because the false vacuum decays too fast.

\section{Conclusion}

To conclude, we discussed some general features of quantum gravity,
such as renormalizability, complementarity, minimal length,
definition of observables, and weak gravity conjecture. We also
considered phenomenological models such as ghost inflation,
non-commutative inflation, brane inflation, k-inflation and resonant
tunneling. All these effects and models show evidence that eternal
inflation is prohibited by quantum gravity.

One should note that a full quantum gravitational treatment for
eternal inflation is not so far available. The arguments and models
considered above are inspired by some features of quantum gravity.
We hope we can gather some information for eternal inflation from
these arguments and models. It may turn out that some of these
models are ruled out by experiments or theoretical developments in
the future. On the other hand, some of these models require that the
scale of quantum gravity is significantly lower than $10^{19}$ GeV,
which needs some luck or fine-tuning. However, since all the above
models are against eternal inflation, we infer that eternal
inflation may be really problematic when quantum gravity effects are
taken into consideration.

\section*{Acknowledgments}
We thank Yi-Fu Cai, Miao Li, Jian-Xin Lu and Yang Zhou for
discussion. We also thank Yi Ling and Nanchang University for
hospitality during the preparation of this work.

\end{document}